\documentclass[epj]{svjour}
\usepackage{graphics}
\newcommand{\text}[1]{{\rm #1}}
\newcommand{\etal}{{\emph{et al.}}}
\begin{document}
\title{Microscopic models for exotic nuclei}
\author{Michael Bender\inst{1}\thanks{Conference presenter} \and 
        Paul-Henri Heenen\inst{2}}
\institute{Physics Division, Argonne National Laboratory,
           Argonne, Illinois 60439, U.S.A. \and
           Service de Physique Nucl{\'e}aire Th{\'e}orique,
             Universit{\'e} Libre de Bruxelles,
             CP 229, B--1050 Brussels, Belgium}

\date{February 18, 2005}
\abstract{
Starting from successful self-consistent mean-field models,
this paper discusses why and how to go beyond the mean field
approximation. To include long-range correlations 
from fluctuations in collective degrees of freedom, one has 
to consider symmetry restoration and configuration mixing, 
which give access to ground-state correlations and spectroscopy.
} 
%
%
\maketitle
\section{Self-Consistent Mean-Field Models}
\label{sec:mf}
Self-consistent mean-field models are one of the standard approaches 
in nuclear structure theory, see Ref.\ \cite{RMP} for a recent
review. For heavy nuclei, they are the only fully microscopic method 
that can be applied systematically. 
%
%
\subsection{Ingredients}
There are three basic ingredients of self-consistent mean-field
models: (see Ref.\ \cite{RMP} for references)

1.) the many-body state is assumed to be an indepen\-dent-quasi-particle 
state of the BCS type. Degrees of freedom are a set of orthonormal 
single-particle states $\phi_k$ with corresponding operators $\hat{a}_k$
and occupation amplitudes $v_k$. The generalized one-body 
density matrix is idempotent
\begin{equation}
\label{eq:R}
\mathcal{R}^2 
= \mathcal{R}
= \left(
  \begin{array}{cc}
  \rho       & \kappa     \\
  -\kappa^*  & 1 - \rho^*  
  \end{array} \right)
= \left(
  \begin{array}{cc}
  \langle \hat{a}^\dagger \hat{a} \rangle         & \langle \hat{a} \hat{a} \rangle     \\
  \langle \hat{a}^\dagger \hat{a}^\dagger \rangle & \langle \hat{a} \hat{a}^\dagger \rangle  
  \end{array} \right)
.
\end{equation}

2.) An effective interaction tailored for the purpose of 
mean-field calculations has to be used. It incorporates the short-range
correlations induced by the strong interaction. 
The actually used effective interactions are parametrized and adjusted 
phenomenologically. They are formulated either as a density-dependent 
two-body force or as an energy functional $\mathcal{E}$ depending on 
the density matrix in the spirit of density functional theory.

3.) The equations-of-motion for the single-particle states and 
the occupation amplitudes are determined self-consis\-tent\-ly from 
the variation of the total energy adding constraints on the particle 
number
\begin{equation}
\label{eq:mf}
\delta \Big( \mathcal{E} 
  - \lambda_N \langle \hat{N} \rangle
  - \lambda_Z \langle \hat{Z} \rangle
  + \cdots
  \Big) 
= 0
.
\end{equation}
This leads to the Hartree-Fock-Bogoliubov (HFB) equations,
or, using a common approximation, the HF+BCS equations.
The Lagrange parameters $\lambda_i$ are adjusted to meet 
conditions for the constraint, for example 
\mbox{$\langle \hat{N} \rangle = N$} for the average neutron number.
The main ingredients of the HFB equations are the 
single-particle Hamiltonian and the pairing field, which 
are obtained as first functional derivatives of the total 
energy
\begin{equation}
\label{eq:sph}
\hat{h} 
= \frac{\delta \mathcal{E}}{\delta \rho} 
,
\qquad
\hat{\Delta} 
= \frac{\delta \mathcal{E}}{\delta \kappa^*}
.
\end{equation}
%
%
\subsection{Typical Applications}
Mean-field models can be used to describe a manifold of 
phenomena and experimental data:

\noindent
$\bullet$ Nuclear masses or binding energies, and all difference 
quantities derived from them, like one- and two-particle-separation 
energies, $Q$ values for $\alpha$ and $\beta$-decay. 

\noindent
$\bullet$ Deformation energy surfaces can be mapped by adding 
one or more constraints on a multipole moment 
\mbox{$ -\lambda_{\ell m} \langle \hat{Q}_{\ell m}\rangle$} 
to the variational equation (\ref{eq:mf}).

\noindent
$\bullet$ the radial density distribution and quantities derived 
from it as the mean-square radii of the charge and neutron 
distributions, the neutron skin, the surface thickness, or the full 
charge form factor at low momentum transfer.

\noindent
$\bullet$ the spatial density distribution, for example
multipole moments of well-deformed nuclei.

\noindent
$\bullet$ 
The very concept of a single-particle energy, associated with the
eigenvalues of the single-particle Hamiltonian $\hat{h}$, 
Eq.\ (\ref{eq:sph}), refers to an underlying mean-field picture 
of the nucleus.
Experimental single-particle energies, however, are obtained as
an energy difference between the ground-state of an even-even
nucleus on one the hand and states in adjacent odd-$A$ nuclei
on the other, the latter having a different structure 
due to the unpaired nucleon, which adds significant corrections.

\noindent
$\bullet$ Rotational bands of well-deformed nuclei can be
obtained by adding a constraint on one component of the angular momentum
$-\omega_i \langle \hat{J}_i \rangle$ to the variational equation 
(\ref{eq:mf}), which is equivalent to solving the mean-field equations 
in a rotating frame. This adds inertial forces to the modeling, which
align the angular momenta of the single-particle states and weaken pairing 
with increasing total angular momentum.
%
%
\subsection{Prospects}

\noindent
$\bullet$ Mean-field models offer an intuitive interpretation 
of their results in terms of the shapes of a nuclear  liquid 
and of shells with single-particle states.


\noindent
$\bullet$ The full model space of occupied states
can be used, removing any distinction between core and 
valence particles and the need for effective charges.

\noindent
$\bullet$ This allows the use of a \emph{universal} effective interaction,
universal in the sense that it can be applied for all nuclei throughout 
the periodic chart. There is, however, no consensus among practitioners 
of the field about a \emph{unique} effective interaction. Many 
different functional forms have been proposed -- for example 
non-relativistic Skyrme and Gogny interactions and finite-range as 
well as point-coup\-ling relativistic interactions -- and parameterizations 
there\-of to be found in the literature \cite{RMP}.
%
%
\subsection{Difficulties and Problems}
%
%
\begin{table}[t!]
\begin{tabular}{lll}
\hline\noalign{\smallskip}
symmetry               & generator        &  which states              \\
\noalign{\smallskip}\hline\noalign{\smallskip}
$U(1)$ gauge           & particle number  &  pairing                   \\
translational          & momentum         &  finite nuclei             \\
rotational             & angular momentum &  deformation               \\
parity                 & parity           &  octupole deformation      \\
\noalign{\smallskip}\hline
\end{tabular}
\caption{\label{tab:symmetries}
Examples for symmetries broken in the intrinsic frame of the nucleus.
}
\end{table}
%
%

\noindent
$\bullet$
An independent particle-description establishes a body-fixed  
intrinsic frame of the nucleus. The connection of mean-field
results to spectroscopic observables in the laboratory frame
of reference relies on additional assumptions like the rigid-rotor 
model, which are not valid, for example, at small deformation 
or in soft nuclei. 

\noindent
$\bullet$
By construction, a mean-field state breaks symmetries in the 
laboratory frame. Examples are given in Table~\ref{tab:symmetries}.
On the one hand, symmetry breaking is a desired feature of 
mean-field models. In the language of the spherical 
shell model (using a spherically symmetric Slater
determinant as reference state) the symmetry-breaking in mean-field 
models adds the most important $n$-particle-$n$-hole and 
particle-particle correlations to the modeling at very 
moderate computational cost. On the other hand, a broken symmetry 
mixes excitations related to the symmetry operator into the
mean-field state. For example, broken rotational symmetry
mixes states with different values of $J^2$, i.e.\ the members 
of a rotational band. Broken parity mixes states of opposite
parity, broken translational symmetry admixes states with
different center-of-mass motion. Restoring the symmetries
decomposes the mean-field states into states with proper
quantum numbers.

\noindent
$\bullet$ 
The mean-field approach becomes ill-defined when the
binding energy changes slowly with a collective degree 
of freedom. This is a common situation in transitional nuclei.

\noindent
$\bullet$
It is tempting to associate two or more local minima in the 
potential energy landscape that are separated by a substantial 
barrier with different physical states, so-called shape coexistence.
This interpretation might not always be valid as two different 
mean-field states are not orthogonal, and they might well be 
coupled by the interaction.
%
%
\section{Going beyond the mean field}
\label{sec:bmf}
The idea is to start from self-consistent mean-field models
as described above, keeping their advantages and successes, and 
to resolve the remaining problems in an efficient, systematic 
and consistent manner. This will add long-range correlations to the 
model, where ''long-range'' does not refer to the range of an 
interaction, but to collective correlations that involve the 
nucleus as a whole.

Two kinds of correlations have to be distinguished. As outlined above,
a mean-field state describes static correlations related to deformation 
or pairing. These have to be distinguished from the dynamical 
correlations we will discuss below. They are also related to deformation 
and pairing, but describe fluctuations in collective degrees of freedom.
The dynamical correlations cannot be described by a state for which 
\mbox{$\mathcal{R}^2 = \mathcal{R}$} holds and therefore require to 
go beyond the mean field.
%
%
\subsection{Projection Methods}
\label{subsec:proj}
As a first-order approximation to projection, corrections to the energy 
are used in self-consistent mean-field models. Most prominent examples 
are the center-of-mass correction and the Lipkin-Nogami scheme to calculate
the occupation amplitudes. Both are approximations to projection before 
variation (on zero momentum and particle number, respectively),
when consistently included in the variation, Eq.\ (\ref{eq:mf}).
Sometimes a rotational correction to the binding energy is also applied. 
The corrections work best when the symmetry breaking is large, which is 
often not the case. It is, therefore, desirable to restore broken 
symmetries of the mean-field states exactly by projecting on good 
quantum numbers after variation. The projection might be 
combined with some of the correction schemes.

Eigenstates of the particle-number operator $\hat{N}$, with eigenvalue 
$N_0$, are obtained applying the particle-number projection operator
\begin{equation}
\hat{P}_{N_0}
= \frac{1}{2 \pi}
  \int_{0}^{2 \pi} \! d \phi_N \;
  e^{i \phi_N (\hat{N}-N_0)}
,
\end{equation}
while eigenstates of the angular momentum operators $\hat{J}^2$
and $\hat{J}_z$, with eigenvalues \mbox{$\hbar^2 J (J+1)$} and 
$\hbar M$, are obtained applying the operator
\begin{equation}
\label{eq:PJ}
\hat{P}^J_{MK}
= \frac{2J+1}{16 \pi^2}
  \! \int_{0}^{4\pi} \! \! d\alpha 
  \! \int_0^\pi \! \! d\beta \; \sin(\beta) 
  \!  \int_0^{2 \pi} \! \! d\gamma \;
  \mathcal{D}^{*J}_{MK} \, 
  \hat{R} 
,
\end{equation}
where \mbox{$\hat{R} = e^{-i\alpha \hat{J}_z}
e^{-i\beta \hat{J}_y} e^{-i\gamma \hat{J}_z}$} is the rotation 
operator and $\mathcal{D}^{*J}_{MK}$ a Wigner function. Both depend 
on the Euler angles $\alpha,\beta,\gamma$. $\hat{P}^J_{MK}$
picks the component with angular momentum projection $K$ along the
intrinsic $z$ axis. The projected state is then obtained by 
summing over all $K$ components with weights determined from
a variational equation. Note that $\hat{P}^J_{MK}$ is not a 
projection operator in the strict mathematical sense \cite{Rin80a}. 

\begin{figure}
\resizebox{0.5\textwidth}{!}{%
  \includegraphics{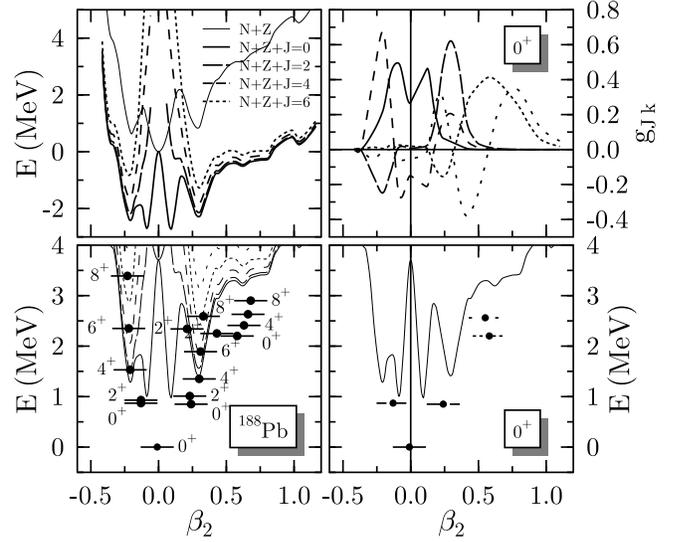}
}
\caption{\label{fig:jproj}
Decomposition of the energy into angular-momentum components 
(upper left), collective wave function (upper right) and 
energy (lower right) for the mixed \mbox{$J=0$} states,
and complete spectrum of low-$J$ states (lower left) for $^{188}$Pb.
All curves are plotted against the mass quadrupole 
deformation $\beta_2$ of the unprojected mean-field states.
}
\end{figure}

In the current implementation of our model, we start with HF+BCS 
or HFB states $|q \rangle$ for which we assume even particle numbers,
good parity, axial and time-reversal symmetry. This allows for 
the analytical evaluation of the $\alpha$ and $\gamma$ integration 
in Eq.\ (\ref{eq:PJ}) at the price of restricting the projected 
states to positive parity $P=+1$, even integer total angular 
momentum $J$, and intrinsic angular momentum projection $K=0$
\begin{equation}
| J M q \rangle
= \frac{\hat{P}^J_{M0} \; \hat{P}_{N_0} \; \hat{P}_{Z_0} |q \rangle}
       {\langle q | \hat{P}^J_{00} \; \hat{P}_{N_0} \; \hat{P}_{Z_0} |q \rangle^{1/2}}
.
\end{equation}
As an example, the upper left panel 
of Fig.\ \ref{fig:jproj} shows the decomposition of the particle-number 
projected energy curve for $^{188}$Pb (thin solid line) into its 
angular-momentum  components (thick curves). All energies are normalized 
to the spherical state. The intrinsic spherical state is a pure \mbox{$J=0$} 
state by construction. The difference between the mean-field and projected 
\mbox{$J=0$} states is the rotational energy. It increases rapidly to about 
3 MeV at small $\beta_2$, and then grows at a slower rate with deformation. 

The example of $^{188}$Pb demonstrates that projection after 
variation of the mean-field ground-state might not lead to the 
lowest projected state as it is not a variational procedure. Instead, 
one has to consider a set of states with different deformations and 
search for the energy minimum. In particular, the lowest projected state of 
a nucleus with a spherical mean-field ground-state is usually obtained 
from a deformed state. For such nuclei, there is the additional 
peculiarity that one obtains two \mbox{$J=0$} minima at small oblate 
and prolate deformation, see Fig.\ \ref{fig:jproj}. Closer 
examination reveals that they represent the same state, as their 
overlap is very close to one.
%
%
%
%
%
\subsection{Variational Configuration-Mixing}
\label{subsec:gcm}
The ambiguities of many near-degenerated states with different 
deformation can be overcome by diagonalizing the Hamiltonian in 
the space of these states within the Generator Coordinate Method
(GCM). The mixed projected many-body state is set-up as a
coherent superposition of projected mean-field states
$| J M q \rangle$ with different intrinsic deformations $q$
\begin{equation}
| J M k \rangle
= \sum_{q} f_{J k} (q) \, | J M q \rangle
,
\end{equation}
where $f_{J,k} (q)$ is a weight function which
is determined from the stationarity of the states
\begin{equation}
\frac{\delta}{\delta f_{J k}^*}
\frac{\langle J M k | \hat{H} | J M k \rangle}
     {\langle J M k | J M k \rangle}
= 0
,
\end{equation}
which leads to the Hill-Wheeler-Griffin equation \cite{Hil53a}
\begin{equation}
\label{eq:HWG}
\sum_{q'}
\big[ \langle J M q | \hat{H} | J M q' \rangle 
      - E_k \, \langle J M q | J M q' \rangle \big] \; f_{J,k} (q') 
= 0
,
\end{equation}
that gives a correlated ground state for each value of $J$, and, in 
addition, a spectrum of excited states. The weight functions 
$f_{J k} (q)$ are not orthonormal. A set of orthonormal collective 
wave functions $g_{J k} (q) = \langle J M k | q \rangle$ in the basis of
the intrinsic states is obtained from a transformation involving the 
square root of the norm kernel.

The actual choice for the generator coordinate depends on the mode to be 
described, for example, the quadrupole or octupole moment of the mass 
density, or the monopole moment of the pair density, which then delivers 
a description of quadrupole, octupole or pairing vibrations, respectively. 
Examples for such calculations, without angular-momentum projection, can 
be found in Ref.\ \cite{Hee01a}. Several generator coordinates can be easily
combined for multi-dimensional calculations, although this has been rarely 
done so far. For 
all results shown  here, the axial quadrupole moment of the mass distribution
serves as the generator coordinate. Hence, the excited states are 
either quadrupole vibrational or rotational states. The right panels of 
Fig.\ \ref{fig:jproj} show as an example the mixing of the \mbox{$J=0$} 
states with different quadrupole moments in $^{188}$Pb. The upper right 
panel shows the collective wave functions $g_{0 k} (q)$ for the five lowest 
collective \mbox{$J=0$} states, the lower right  panel the corresponding 
energies drawn in the same line style by horizontal lines centered at the
average deformation of the mean-field states they are composed of
together with the projected \mbox{$J=0$} energy curve. The energies are 
now normalized to the projected GCM ground state. Combining such calculations
for all values of $J$ gives then the the entire spectrum of low-$J$ states, 
most of which can be clearly grouped into rotational bands, see the 
lower left panel of Fig.\ \ref{fig:jproj}, 

The projected energy curves in Fig.\ \ref{fig:jproj} are the diagonal
matrix elements entering the Hill-Wheeler equation. They should not be
confused with a collective potential, which does not exist in 
the GCM framework -- nor does a collective mass. Both appear in 
approximations like the Bohr-Hamiltonian \cite{Bru,Fle04a}. Owing to
the energy gain from configuration mixing, the GCM ground state is 
located below the energy curves.
%
%
\begin{figure}
\resizebox{0.5\textwidth}{!}{%
  \includegraphics{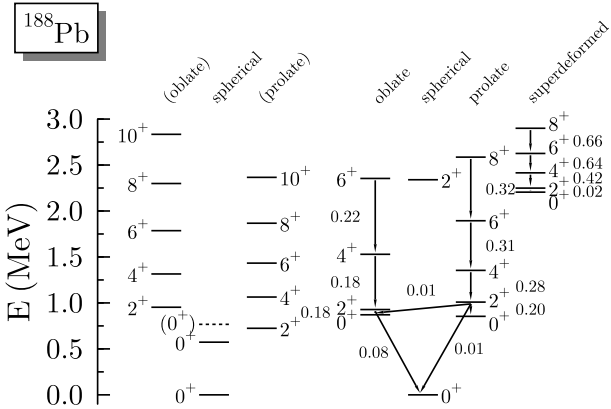}
}
\caption{\label{fig:spectrum}
Experimental (left) and calculated (right)
excitation spectra and selected transition quadrupole
moments $\beta_2^{(t)}(J_{k'}' \to J_k)$ for $^{188}$Pb.
}
\end{figure}
%
%

Projection is a special case of the GCM, where exactly degenerate
states are mixed. The generators of the group involved define 
the collective path, and the weight functions are determined 
by the restored symmetry. Angular-momentum projection 
is part of the quadrupole correlations, as 
it mixes states with different orientations of the quadrupole tensor. 
Therefore the GCM mixing of states with respect to the quadrupole moment 
should be performed together with angular-momentum projection.

For a state resulting from the mixing of different mean-field states, 
the mean particle number is not anymore equal to the particle number 
of the original mean-field states. Projection, as done here, eliminates 
this problem, otherwise a constraint on the particle number has to be 
added to the Hill-Wheeler-Griffin equation (\ref{eq:HWG}).
We also perform an approximate particle-number projection before 
variation in the Lipkin-Nogami approach to ensure that 
pairing correlations are present in all mean-field states.


The angular-momentum projected GCM  allows to calculate transition 
moments directly in the lab frame for in-band and out-of-band 
transitions, for example $B(E2)$ values
\begin{eqnarray}
\lefteqn{
B (E2; J_{k'}' \to J_k) }\\
& = & \frac{e^2}{2J'+1}\sum_{M =-J }^{+J }\sum_{M'=-J'}^{+J'}\sum_{\mu=-2}^{+2}
      | \langle J M k | \hat{Q}_{2 \mu} | J' M' k' \rangle |^2
\nonumber 
.
\end{eqnarray}
The $B(E2)$ value scales with mass and angular momentum. A more
intuitive measure is the transitional quadrupole deformation obtained
from the $B(E2)$ using the static rotor model
\begin{equation}
\beta_2^{(t)}(J_{k'}' \to J_k)
= \frac{4 \pi}{3 R^2 A}
  \sqrt{
  \frac{B(E2;J_k \to J_k'-2)}
       {( J \,0 \, 2 \, 0 | J-2 \, 0 )^2 e^2}}
\end{equation}
with $R=1.2 \, A^{1/3}$. The method gives also the spectroscopic 
quadrupole moments in the lab frame
\begin{equation}
Q_{s}(J_k)
=  \langle J \; M\!\!=\!\!J \; k | {\hat Q}_{20} | J \; M\!\!=\!\!J \; k \rangle
\end{equation}
which again scale with mass and angular momentum and might be
given in a more intuitive measure through a dimensionless
deformation parameter
\begin{equation}
\beta_2^{(s)}(J_k)
= - \frac{2J+3}{J} \,
   \sqrt{\frac{5}{16 \pi}} \, \frac{4 \pi}{3 R^2 A} \, Q_{s}(J_k)
\end{equation}
again with $R=1.2 \, A^{1/3}$ and assuming axial symmetry. 
For a given rotational band in the rigid 
rotor, one has $\beta_2^{(t)}(J+2 \to J) \approx \beta_2^{(s)}(J+2)
\approx \beta_2^{(s)}(J)$ for \mbox{$J>0$}. Deviations from this behavior 
point at a more complicated situation. An example is given in 
Fig.\ \ref{fig:spectrum}. For high-$J$ states, the $\beta_2^{(t)}$
values within a band are constant, while with the mixing of the 
low-lying states they are significantly decreased.
%
%
\section{Examples of Applications}
All results discussed here were obtained by configuration mixing
of particle-number and angular-momentum-pro\-jec\-ted HF+BCS states
with different axial mass quadrupo\-le moments. 
We chose the Skyrme interactions SLy4 or SLy6 \cite{SLyx} for the 
particle-hole channel and a density-de\-pen\-dent delta pairing 
interaction (``surface pairing'') \cite{Rig99a} for the 
particle-particle channel. A group in Madrid uses the Gogny force 
in a similar model \cite{Madrid}.
%
%
\subsection{Spectroscopy}
Results for spectroscopic observables obtained with our method for $^{24}$Mg 
have been published in \cite{Val00a}, for $^{16}$O in \cite{Ben03a}, 
for $^{32}$S, $^{36,38}$Ar, $^{40}$Ca in \cite{Ben03b}, for $^{186}$Pb in 
\cite{Dug03a}, for  $^{182-194}$Pb in \cite{Ben04b}, and for $^{240}$Pu 
in \cite{Ben04c}. 
%
%
\begin{figure}[t!]
\resizebox{0.5\textwidth}{!}{%
  \includegraphics{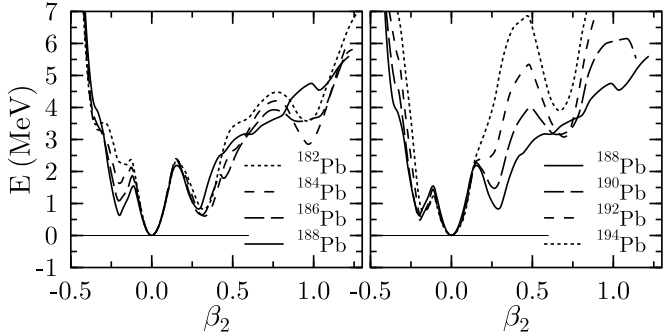}
}
\caption{Mean-field deformation energy curves for Pb isotopes.}
\label{fig:pbsyst1}
\end{figure}
%
%

\begin{figure}[t!]
\resizebox{0.5\textwidth}{!}{%
  \includegraphics{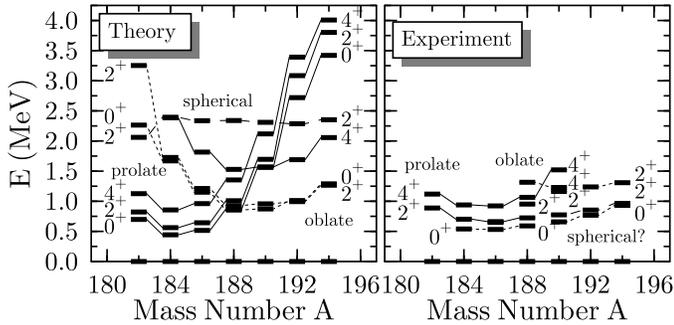}
}
\caption{Lowest collective states in the Pb isotopes.}
\label{fig:pbsyst2}
\end{figure}

The neutron-deficient Pb isotopes show unique spectroscopic features,
which are associated with a spherical ground state, an oblate minimum 
present above \mbox{$A=188$} but disappearing below, a prolate 
minimum present below \mbox{$A=188$} and disappearing above, and 
a superdeformed minimum, that is confirmed down to
\mbox{$A=192$}, see Fig.\ \ref{fig:pbsyst1}. Projected GCM then delivers 
collective states that can be associated with a spherical ground state 
and excited prolate, oblate and superdeformed bands. There is a nice 
qualitative agreement with experimental data, see Fig.\ \ref{fig:pbsyst2}, 
but the calculated transition energies within the bands are 
too dilute, see also Fig.\ \ref{fig:spectrum} for $^{188}$Pb. 
For more details and further discussion of other observables, 
see Refs.\ \cite{Dug03a,Ben04b}.
%
%
%
%

An example with very different spectroscopic features is the well-deformed 
nucleus $^{240}$Pu, see Ref.\ \cite{Ben04c}. Projection does not alter the 
topology of the potential energy curve, but gives about 3 MeV 
additional binding for the ground state and about 4 MeV for the 
fission isomer which has now an excitation energy that is 1 MeV lower
compared to mean-field calculations. Projection lowers the outer 
barrier as much as 2 MeV. GCM does not substantially mix states.
Again, the excitation energies within 
the rotational bands are too large, while the deformation is well
described on all levels of approximation: we obtain \mbox{$\beta_2  = 0.29$} 
for the mean-field ground state, \mbox{$\beta_2^{(t)}(J+2 \to J) = 0.30$} 
for all $E2$ transitions within the ground-state band, where all 
excited states have spectroscopic quadrupole moments corresponding 
to $\beta_2^{(s)}(J) = 0.30$, in agreement with the 
experimental value of 0.29 deduced from the \mbox{$B(E2; 0^+_1 \to 2^+_1)$}.
%
%
\subsection{Mass Systematics}

Mass formulas based on self-consistent mean-field models using 
Skyrme interactions have reached a quality where they compete with the 
best available microscopic-macros\-co\-pic models \cite{Ton00a}.
A key to this success is to add various correlation energies 
phenomenologically through correction terms, as a Wigner energy term 
or a rotational correction. There is no correction for vibrations, as
it cannot be formulated in terms of a simple expression. Our model
allows us to consistently calculate the quadrupole correlation energy 
from symmetry restoration and fluctuations of the quadrupole moment.
%
%
\begin{figure}[b!]
\resizebox{0.5\textwidth}{!}{%
  \includegraphics{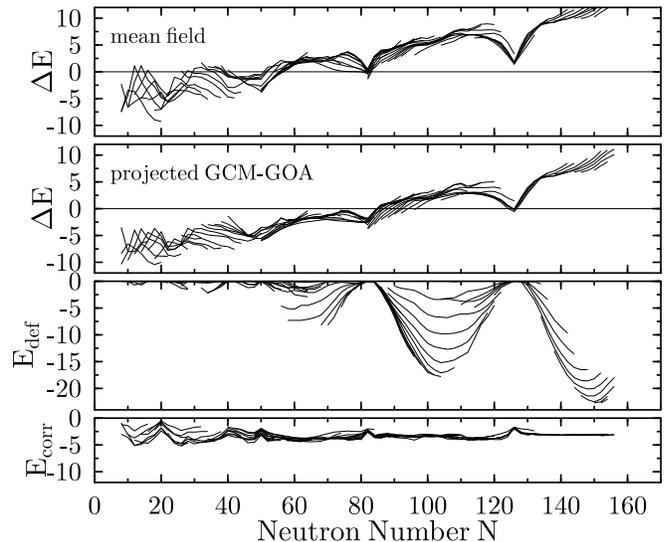}
}
\caption{Upper two panels: difference between calculated and experimental 
masses. Lower two panels: mean-field deformation energy $E_{\rm def}$ and 
beyond-mean-field quadrupole correlation energy $E_{\rm corr}$.
All panels share the same energy scale in MeV. 
}
\label{fig:masses}
\end{figure}
%
%

For the calculation of a mass table including correlations, it 
was necessary to use an approximation to the method described above.
For this purpose, we implemented the Gaussian overlap approximation (GOA) 
into our me\-thod \cite{Ben04a}. While most applications of the GOA 
use it as an intermediate step to derive a Bohr-Hamiltonian
\cite{Bru,Fle04a}, we use the GOA solely as a numerical tool: a 
topological GOA to estimate the integrals over Euler angles from two 
or three exactly calculated points, and a second GOA to
construct the matrices entering the Hill-Wheeler equation from
diagonal matrix elements and matrix elements between nearest neighbors 
only. The GOA puts emphasis on the correlated $0^+$ ground state; most 
information for spectroscopy is lost. Particle-number projection 
is still performed exactly. 
The accuracy of the GOA is better than 300 keV, which is 
sufficient for a study of the systematics of quadrupole correlation 
energies, which are an order of magnitude larger. 

\begin{figure}[t]
\begin{center}
\resizebox{8cm}{!}{
  \includegraphics{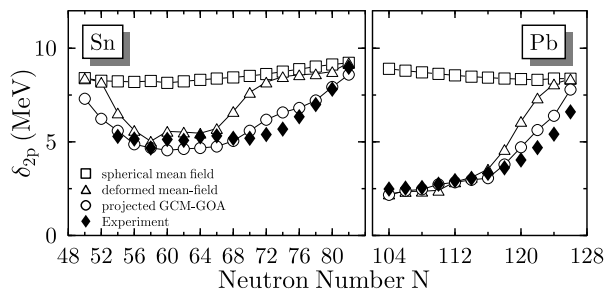}
}
\end{center}
\caption{Two-proton gap $\delta_{2p}$ for Sn and Pb isotopes.
}
\label{fig:d2p}
\end{figure}

Figure \ref{fig:masses} shows some results \cite{Ben04pp}. The 
overall erroneous trend with $A$, that was already observed in
Ref.\ \cite{Sto03a}, can be removed with a slight refit of the coupling 
constants of SLy4 on the mean-field level \cite{Ber05a}. The
quadrupole correlation energy improves the masses by reducing 
the oscillations between the closed shells, but without removing 
them completely. Still, there is a substantial improvement, which 
becomes obvious when looking at energy differences like the 
two-proton gap $\delta_{2p}(N,Z) = E (N,Z-2) - 2 E(N,Z) + E(N,Z+2)$, 
see Fig.\ \ref{fig:d2p}. Spherical mean-field calculations 
(open squares) give near-constant $\delta_{2p}$ in accordance with the stable
\mbox{$Z=82$} shell. Allowing for deformation (open triangles)
and adding correlations (open circles) substantially reduces $\delta_{2p}$ 
for mid-shell nuclei far from $^{208}$Pb, in agreement with experiment
(full diamonds). Similar results were obtained in Ref.\ \cite{Fle04b}.

The lower panels of figure \ref{fig:masses} compare mean-field 
deformation and beyond-mean-field quadrupole correlation energies.
While heavy nuclei are dominated by the deformation energy, light
nuclei are dominated by the correlation energy.
%
%
%
%
%
%
\section{Summary and Outlook}
Projection and configuration mixing significantly improve the modeling 
of nuclei in self-consistent mean-field approaches and give access 
to spectroscopy. Masses are significantly improved around closed shells, 
and the overall structure of collective bands is reproduced, even in 
complicated systems like neutron-deficient Pb isotopes where many 
structures coexist. On the quantitative level, 
neither masses nor excitation spectra are yet described with the desired
precision. This might be for a number of origins. There will be 
imperfections of the effective interactions that we use. Some aspects of 
the effective interaction might be much more sensitive to spectroscopy 
than to the ground states they are fitted to. For consistency, 
the effective interaction should be refitted including the correlations.
To obtain a more robust extrapolation of the interaction into the unknown, 
it is desirable to establish a link between the effective interaction 
needed for calculations as done here, and more \emph{ab-initio} methods.

On the other hand, the modeling of the configuration mixing might still
have some deficiencies as well. There are additional modes like pairing 
vibrations, triaxial quadrupole deformations, or octupole vibrations, which 
might play a role for certain nuclei and, therefore, should be 
included in a unified model. The determination of the collective path has to
be re-examined, and diabatic states may play a role in some situations.
An interesting insight comes from self-consistent, cranked mean-field 
calculations: for $^{240}$Pu, the excitation energies from cranked HFB are 
in much better agreement with experiment than our projected values when 
using the same interaction \cite{Ben04c}. Cranked mean-field states break 
time-reversal invariance and have the proper angular momentum on the 
average, which might be crucial for excitation energies. A generali\-zation 
of our model to use cranked states as a starting point for projected GCM 
is currently underway. This will also allow to describe nuclei with an odd 
nucleon number in our framework. A lot of work is left for the future, 
but present results are most encouraging.
%
%
\section*{Acknowledgements}
The results discussed here were obtained in collaboration 
with G.\ F.\ Bertsch, P.\ Bonche, T.\ Duguet, and H.\ Flocard.
We thank T.\ Duguet and R.\ V.\ F.\ Janssens for critical 
reading of the manuscript.
MB thanks for the warm hospitality at the Service de 
Physique Nucl{\'e}aire Th{\'e}orique at the Universit{\'e} 
Libre de Bruxelles, Belgium, and the Institute for Nuclear Theory,
Seattle, USA, where parts of the research 
presented here were carried out. This work was supported in parts 
by the U.S.\ Department of Energy, Office of Nuclear Physics, 
under Grant W-31-109-ENG-38 (Argonne National Laboratory) and 
the Belgian Science Policy Office under contract PAI P5-07.
%
%

\end{document}